# Generating high-dimensional chaotic signals by the sum

A.A.Kipchatov and L.V.Krasichkov

*College of Applied Science, Saratov State University,*
*83 Astrakhanskaya, Saratov 410071, Russian Federation*

## Abstract

The method of generating of high-dimensional oscillations on the basis of summing of low-dimensional chaotic signals of noncoupled dynamical systems is investigated. It is shown that the correlation dimension of attractor reconstructed from such oscillations is equal to the sum of dimensions of the original signals. The possibilities for obtaining of homogeneous high-dimensional chaotic attractor are numerically demonstrated with the original signals of Rössler system.

# 1. Introduction

Revealing and understanding of regularities of appearance and of evolution of complex and chaotic behaviour of dynamical systems is connected with the development of analysis methods of their phase space trajectories. In most of experimental and in some of computer simulated dynamical systems the scalar time series of only one of system's variables are accessible [1]. In this case, reconstruction of attractor by Takens' time delay method [2] and algorithm of correlation dimension estimation [3] are mostly used to characterize the dynamics of such systems.

At present, high-dimensional systems of different nature are widely investigated by these methods, in particular, experimental signals containing a noise or signals of extended systems. For such systems, it is especially important to select the parameters of the correlation dimension algorithm (the delay time, the number of points, the embedding dimension). Attempts to improve the precision of data and to increase the length of time series leads to the extremal increasing of calculations. In order to avoid such effects the determining of appropriate parameter values of the reconstruction as well as of the correlation dimension algorithm is needed [1]. For this purpose, test signals with given dimensions are used.

Oscillations of the well-known systems [1, 3] or quasiperiodic oscillations with an irrational frequency ratio and with some integer dimension [4, 5] are usually used as the test signals. Oscillations generated by more complicated systems (e.g. delay differential equations [4, 6] or lattice [5]) have nonhomogeneous attractors, i.e. it is very difficult to select a linear slope region on the correlation integral plot. Thus, using of these oscillations as test ones could become problematic.

Obtaining of quasiperiodic signals with any given integer dimension is sufficiently difficult in numerical as well as in natural experiments since the synchronization of the large number of oscillators is possible. Furthermore such signals have the discrete spectra and require to set proper values of parameters (the frequency ratio, the amplitude ratio, the sampling time, the length of time series and the data precision) even in the numerical simulations.

Another way of confirmation of reliability of the correlation dimension estimation, suggested recently, is based on using of a surrogate data [7]. Such data are generated from original time series and are similar, in some properties (mean, variance, power spectrum and autocorrelation), to original time series but their dimension is close to dimension of noise. If dimension estimated from surrogate is larger than the dimension of original oscillation then the result of estimation is considered as reliable.

In any case, in order to select the parameters of correlation dimension algorithm appropriately the comparison of dimensions estimated for the data and for the test signals is required.

In this Letter, we present one of the methods of obtaining of test high-dimensional chaotic signals. These signals have the *limited preknown* value of dimension unlike surrogate ones.

## 2. Idea of method

The method of generating of high-dimensional chaotic oscillations is based on increasing of dimension when periodic oscillations are mixed. However, in this method, low-dimensional chaotic oscillations are used as summands. It can be supposed that the dimension of the sum of noncorrelated chaotic oscillations is equal to the sum of dimension of original ones. Similar idea has been just mentioned in Ref. [7]. The chaotic oscillations possess properties, namely, decay of correlations, which allow to avoid the synchronization between the original oscillations and to avoid decreasing of resulted dimension. However, the suggestion that the dimensions of noncorrelated chaotic oscillations are summing under summing these oscillations is not obvious, especially when the attractors of original oscillations have multifractal properties [8]. A rigorous proof of above statement is difficult since the correlation dimension is defined algorithmically. For this reason, we confirm this suggestion numerically in the Letter.

## 3. Methods of analysis

In our experiments the oscillation processes presented by scalar time series are considered. In order to reconstruct system behaviour in the phase space we use Takens' time delay method [2]. On the base of this method the phase space behaviour of the system can be reconstructed from a scalar time series $u(t)$, where $t = nT_S$ ($n = 0, 1, ..., N$), $T_S$ is the sampling time and $N$ is the number of data points. Then state vector in the $m$-dimensional embedding space is given by

$$\mathbf{u}_n \equiv \mathbf{u}(t) = \{u(t), u(t+\tau), ..., u(t+\tau(m-1))\}, \tag{1}$$

where $\tau$ is the delay time which is a multiple of $T_S$ ($t = kT_S$ with $k = 0, 1, ...$).

The correlation dimension used in our experiments is calculated by Grassberger-Procaccia algorithm [3] and the correlation integral for $m$-dimensional embedding vectors $u(t)$ is estimated as follows (see, for example, refs. [9, 10])

$$C(\varepsilon) \approx \frac{1}{M} \sum_{i=1}^{M} \frac{1}{N} \sum_{j=1}^{N} \theta(\varepsilon - \parallel \mathbf{u}_i - \mathbf{u}_j \parallel), \tag{2}$$

where $M$ is the number of reference points, $N$ is the number of data points, $\theta$ is the Hevisaide function and the vertical bars denote the norm of the vector. In this work, the maximum norm $\parallel \mathbf{u}_i - \mathbf{u}_j \parallel = \max_{q=0,...,m-1} \mid u_{i+kq} - u_{j+kq} \mid$ is used.

Here, the correlation dimension $D_C(\varepsilon)$ is calculated, from a log-log plot of $C(\varepsilon)$ versus $\varepsilon$, by the following equation

$$D_c(\varepsilon) = \frac{\log(C(\varepsilon)/C(\varepsilon+\delta))}{\log(\varepsilon/(\varepsilon+\delta))},$$

where $\delta$ is the localization range of the correlation function in which the linear least-square fit is performed.

The results of estimation of the correlation dimension $D_C$ versus the length scale $E$ are presented in dB units (i.e. $E = 20 \cdot \log_{10}(\frac{\varepsilon}{\varepsilon_0})$, $\Delta = 20 \cdot \log_{10}(\frac{\delta}{\varepsilon_0})$ where $\varepsilon_0$ is the total extantion of attractor), $\Delta$=–50dB. The length scale $E$ expressed in dB units allows to compare the results of dimension estimation with the standard units of measurements of spectral density and of dynamical oscillational range. Moreover, the plot $D_C(E)$ gives more information about the structure of the attractor for different length scales [11] and allows to estimate its nonhomogeneity [9].

To verify the idea of summing of dimensions, it is best to use the time series of well-studied systems as each of summands. From this reason, in our experiments we considered the oscillations of the set of noncoupled Rössler systems [12] described by the following equations

$$\begin{aligned} \dot{x}_i &= -(y_i + z_i), \\ \dot{y}_i &= x_i + e_i y_i, \\ \dot{z}_i &= f_i - \mu_i z_i + x_i z_i, \end{aligned} \qquad (3)$$

where $e_i$, $f_i$ and $\mu_i$ are parameters. The resulted signal is obtained by linear transformation as

$$S = \sum_{i=1}^{L} a_i x_i, \qquad (4)$$

where $L$, $a_i$ and $x_i$ are the number of signals in the sum, the constant coefficients defining the amplitude of variable and the variables of eqs.(3), respectively.

To integrate equations of systems (3) the fourth-order Runge-Kutta algorithm with the time step $dt = 0.04$ (the sampling time is $T_S = 10dt = 0.4$) is used. The first 20000 iterates were discarded as transients for all results. We fix $e_1 = f_1 = e_2 = f_2 = 0.2$, unless otherwise indicated. Throughout this Letter, the embedding dimension $m$ is selected to be equal to the total number of the first-order ordinary differential equations in (3) and (4). Such selection of $m$ ($m \geq D_C$) is quite sufficient for computations [13].

## 4. Numerical tests

Consider three examples of summing of different low-dimensional chaotic oscillations. Let us present the results of calculations for the original Rössler system (3) in figs. 1a and 2a for $L = 1$, $a_1 = 1.0$, $\mu_1 = 4.6$ (the chaotic attractor) for further comparisons. Presented plots illustrate the simple structure of the Rössler attractor which has high homogeneity for wide range of the length scales ($D_C(E) \simeq 2$ for –50dB$\leq E \leq$–20dB).

At first we consider the sum of chaotic and periodic signals (figs. 1b and 2b) for parameters values $L = 2$, $a_1 = a_2 = 1.0$, $\mu_1 = 4.6$ (the chaotic attractor), $\mu_2 = 2.6$ (the limit cycle). One can see that the attractor's structure of such signal is principally keeping for large scales (see the phase portrait in fig. 1b). However, the Poincaré section (fig. 2b) shows that small scale structure of the attractor is complicated, i.e. the attractor becomes thicker. This is well illustrated by correlation dimension $D_C(E)$ which, in range –50dB$\leq E \leq$–20dB, is close to two for the original chaotic attractor (fig. 2a); is close to one for the periodic attractor; and for the sum of these signals is about three (fig. 2b). Thus, the resulted dimension of the sum of the signals is equal to the sum of dimensions of the original signals in this case.

Note that the resulted dimension of the sum of original signals of different amplitudes ($a_1 \neq a_2$) essentially depends on the length scale $E$. So for $a_2 \ll a_1$, the dimension increases on small length scales only, i.e. attractor nonhomogeneity is increased. The area of the function $D_C(E)$ where resulted dimension is equal to the sum of original signals is shifted into range of large $E$ with increasing $a_2$ (fig. 3). Thus, the nonhomogeneity of attractor from the sum of signals can be controlled by the varying of amplitudes of summands in eq. (4).

The second experiment was performed for two chaotic signals (figs. 1c and 2c), in this case $L = 2$, $a_1 = a_2 = 1.0$, $\mu_1 = 4.6$, $\mu_2 = 5.7$. Notice that quantitative and qualitative characteristics of signal for parameter value $\mu = 5.7$ are similar to ones for $\mu = 4.6$, although attractors are slightly different in size. For the sum of such two chaotic signals the attractor is complicated significantly; and in this case its Poincaré section (fig. 2c) is becoming more uniform and similar to one for high-dimensional system [5]. Plot of the correlation dimension (fig. 2c) shows that $D_C(E)$ is close to four in ranges $-50\text{dB} \leq E \leq -20\text{dB}$.

Finally, consider the result of dimension estimation $D_C(E)$ of the sum of three chaotic signals for the following parameters of eqs. (3) and (4): $L = 3$, $a_1 = a_2 = a_3 = 1.0$, $\mu_1 = 4.6$, $\mu_2 = 5.7$, $e_3 = 0.2$, $f_3 = 0.4$, $\mu_3 = 5.7$ (fig. 4). The characteristics of Rössler system with $i = 3$ are similar to ones shown in fig. 1a and 2a. It is seen that the dimension of sum signals is close to sum of dimensions of each signals in range $-40\text{dB} \leq E \leq -20\text{dB}$. Some narrowing of the range is connected with lack of points described attractor on small length scales, even for more large number of points for correlaton integral estimation.

# 5.Conclusion

The results presented above demonstrate that the hypothesis on summing of dimensions under the summing of noncorrelated oscillations is supported. This result provides the simple way to create test signals with a *preknown* (high enough) dimension and with the probability characteristics which are similar to low-dimensional oscillations (see spectra which are not changed under summing in fig.1).

However, the statement on summing of dimensions is correct in the case when the correlation dimension is considered as function of the length scale $E$ for common normalization by $\varepsilon_0$ for all functions $D_C(E)$ using in the sum. Under this normalization the varying of amplitudes $a_i$ of the original oscillations leads to displacing $D_C(E)$ from the length scale $E$ and, consequently, to the nonhomogeneous resulted attractor. Thus, varying $a_i$ gives possibility to control nonhomogeneity of attractor and to provide the required $D_C(E)$.

# Acknowledgements

We would like to thank Professor D.I.Trubetskov for useful discussions. This work was supported by The Russian Fund of Fundamental Research under Grant No. 93-02-16171.

# Figure captions

Fig.1. Reconstructed phase portraits (left column) and power spectra (right column) are shown for signals of Rössler system with $L = 1$, $\mu_1 = 4.6$ (a), for the sum of signals with $\mu_1 = 4.6$, $\mu_2 = 2.6$ (b) and for the sum of signals with the $\mu_1 = 4.6$, $\mu_2 = 5.7$ (c).

Fig.2. Poincaré sections (left column) and correlation dimension $D_C(E)$ (right column) ($N = 2 \cdot 10^4, M = 10^3, m = 6, \tau = 1.2$) are shown for signals of Rössler system with $L = 1$, $\mu_1 = 4.6$ (a), for the sum of signals with $\mu_1 = 4.6$, $\mu_2 = 2.6$ (b) and for the sum of signals with the $\mu_1 = 4.6$, $\mu_2 = 5.7$ (c).

Fig.3. The correlation dimension functions $D_C(E)$ ($N = 2 \cdot 10^4, M = 10^3, m = 6, \tau = 1.2$) versus $a_2$ for parameters values $L = 2$, $a_1 = 1.0$, $\mu_1 = 4.6$, $\mu_2 = 4.6$. Different initial conditions were used here to decorrelate the signals of such systems.

Fig.4. The correlation dimension $D_C(E)$ ($N = 2 \cdot 10^5, M = 10^3, m = 9, \tau = 1.2$) of the sum of three signals for $L = 3$, $a_1 = a_2 = a_3 = 1.0$, $\mu_1 = 4.6$, $\mu_2 = 5.7$, $e_3 = 0.2$, $f_3 = 0.4$, $\mu_3 = 5.7$.

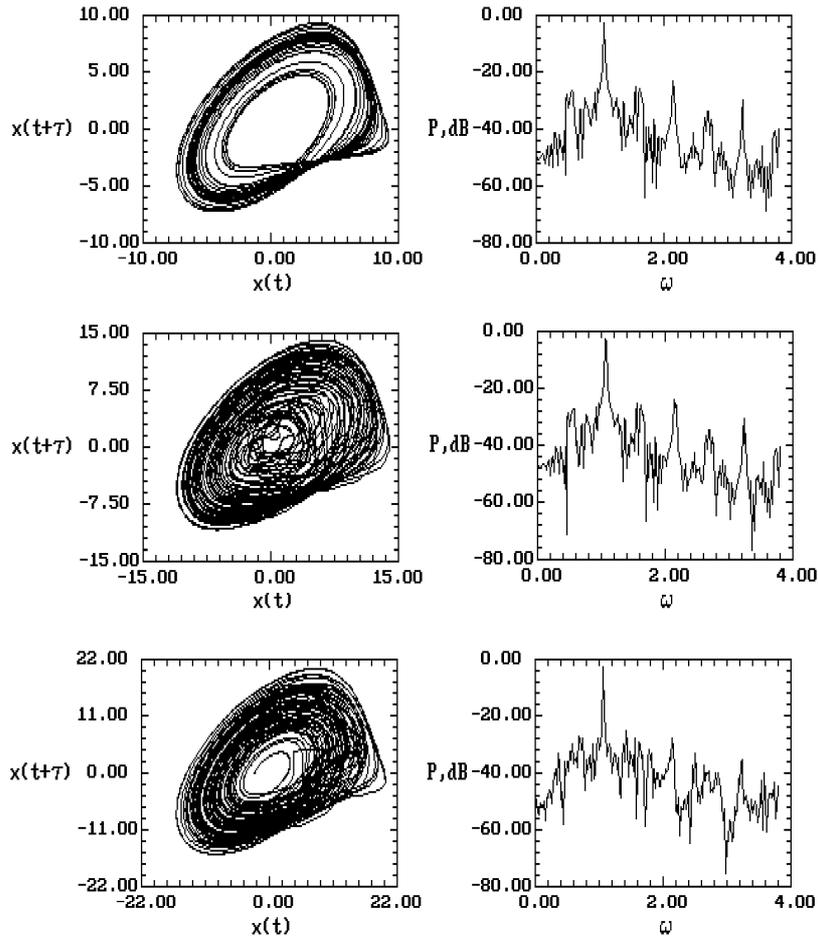

Fig.1. A.A.Kipchatov and L.V.Krasichkov.

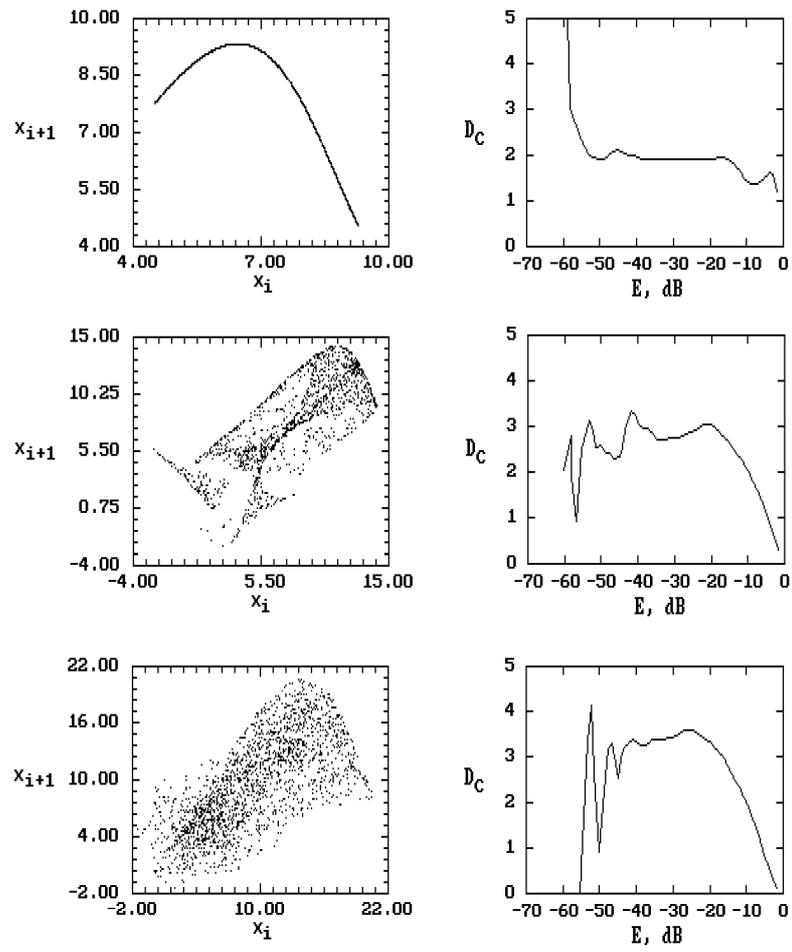

Fig.2. A.A.Kipchatov and L.V.Krasichkov.

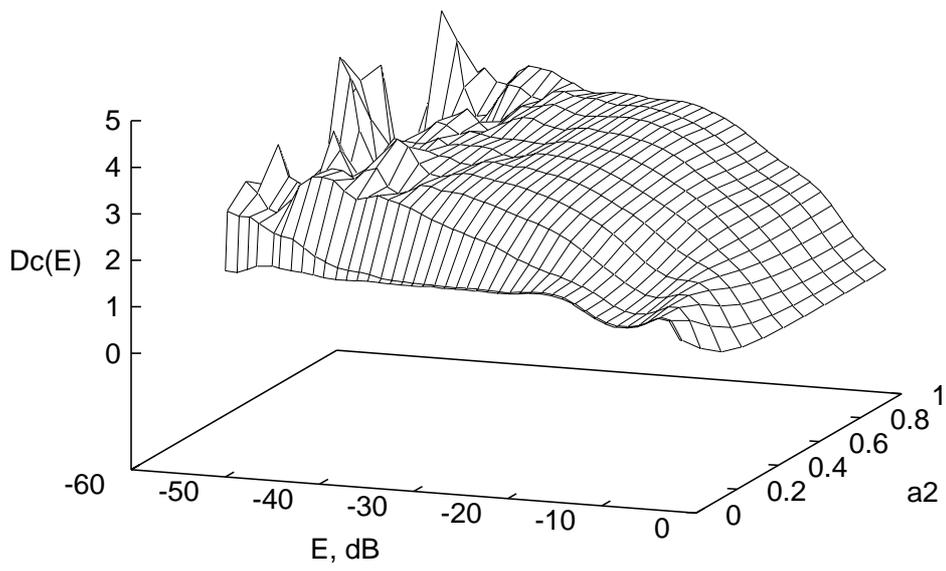

Fig.3. A.A.Kipchatov and L.V.Krasichkov.

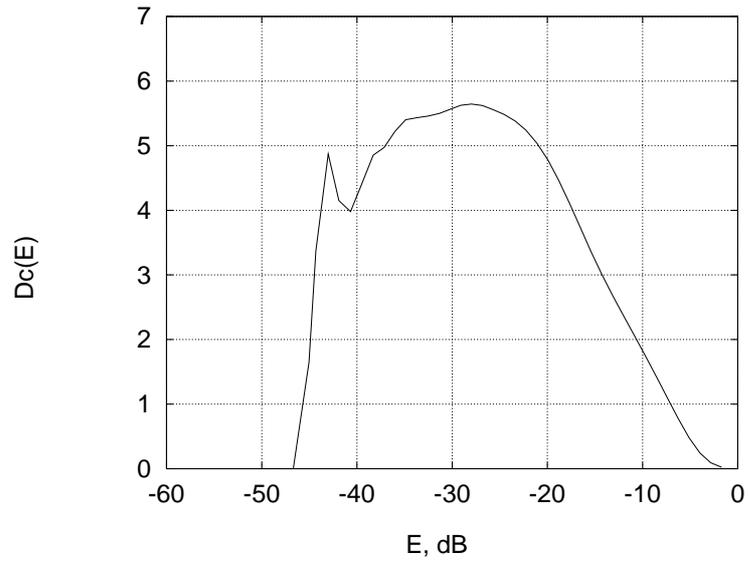

Fig.4. A.A.Kipchatov and L.V.Krasichkov.